\documentclass[12pt]{article}
\usepackage{graphicx}

\begin{document}

\title{New coordinates for the Four-Body problem}

\author{E. Pi\~na \\
Department of Physics \\ Universidad Aut\'onoma Metropolitana - Iztapalapa, \\
P. O. Box 55 534 \\ Mexico, D. F., 09340 Mexico \\
e-mail: pge@xanum.uam.mx}

\date{ }
\maketitle

\abstract{
A new coordinate system is defined for the Four-Body dynamical problem with general masses, having as its origin of coordinates the center of mass. The transformation from the inertial coordinate system involves a combination of a rotation to the principal axis of inertia, followed by three changes of scale leading the principal moments of inertia to yield a body with three equal moments of inertia, and finally a second rotation that leaves unaltered the equal moments of inertia. These three transformations yield a mass-dependent rigid orthogonal tetrahedron of constant volume in the inertial coordinates. Each of those three linear transformations is a function of three coordinates that produce the nine degrees of freedom of the Four-Body problem, in a coordinate system with the center of mass as origin.

The relation between the well known equilateral tetrahedron solution of the gravitational Four-Body problem and the new coordinates is exhibited, and the plane case of central configurations with four different masses is computed numerically in these coordinates.
}
\

{\sl Keywords:} Four-Body Problem. New coordinates.

\

PACS 45.50.Pk Celestial mechanics 95.10.Ce Celestial mechanics(including n-body problems)

\newpage
\section{Introduction}

The coordinate system introduced in this paper is a generalization of the symmetric coordinate system of Pi\~na and Jim\'enez  \cite{pi}, \cite{pj}, \cite{jp}, that was defined for the Three-Body problem. Relative symmetric coordinates in the Three-Body problem were defined by Murnaghan \cite{mu} and Lema\^\i tre \cite{le}. More recently, Hsiang and coworkers have studied at least since 1995 the triangle geometry of this problem \cite{hs}, with an important impact in the modern Three-Body problem reviewed by Chenciner \cite{ch} who posted in the {\sl web} an important panorama on the subject, including the geometry of the so called shape sphere, that is almost the same coordinate system than ours for the case of three particles. Important contributions have been also made by Littlejohn and Reinsch \cite{li} for the analysis of coordinate systems of three and four particles.

The proposal of new coordinates, presented in this paper, have important points of contact with those works, although it sets itself apart from them, and simplifies their ideas in the case of four particles.

\section{The New Coordinates}
The masses of the four bodies $m_1$, $m_2$, $m_3$ and $m_4$ are generally different, and we consider them ordered by the inequalities $m_1 > m_2 > m_3 > m_4$.

We transform from the inertial referential, to the frame of principal axes of inertia by means of a three dimensional rotation $\bf G$ that is parameterized by three coordinates, such as the Euler angles.

In addition to this rotation three more coordinates are introduced, as scale factors $R_1$, $R_2$, $R_3$, where $R_1$, $R_2$ and $R_3$ are three distances closely related to the three principal inertia moments through
\begin{equation}
I_1 = \mu (R_2^2 + R_3^2)\, , \quad I_2 = \mu (R_3^2 + R_1^2) \, , \quad \mbox{and} \quad I_3 = \mu (R_1^2 + R_2^2) \, ,
\end{equation}
where $\mu$ is the mass
\begin{equation}
\mu = \sqrt[3]{\frac{m_1 \, m_2\, m_3\, m_4}{m_1 + m_2 + m_3 + m_4}} \, .
\end{equation}

The size of the scale factors is given in terms of the principal moments of inertia by the equations
\begin{equation}
R_1^2 = \frac{I_2 + I_3 - I_1}{2 \mu}\, , \quad R_2^2 = \frac{I_3 + I_1 - I_2}{2 \mu}\, , \quad R_3^2 = \frac{I_1 + I_2 - I_3}{2 \mu}\, .
\end{equation}

With the first rotation and the change of scale the resulting four body configuration has a moment of inertia tensor with the three principal moments of inertia equal. A second rotation $\bf G'$ does not change this property.

The cartesian coordinates of the four particles, with the center of gravity at the origin, written in terms of the new coordinates are
\begin{equation}
\left(\begin{array}{cccc}
x_1 & x_2 & x_3 & x_4 \\
y_1 & y_2 & y_3 & y_4 \\
z_1 & z_2 & z_3 & z_4
\end{array} \right) = {\bf G} \left(\begin{array}{ccc}
R_1 & 0 & 0 \\
0 & R_2 & 0 \\
0 & 0 & R_3
\end{array} \right) {\bf G'}^{\rm T}
\left(\begin{array}{cccc}
a_1 & a_2 & a_3 & a_4 \\
b_1 & b_2 & b_3 & b_4 \\
c_1 & c_2 & c_3 & c_4
\end{array} \right)\, ,
\end{equation}
where $\bf G$ and $\bf G'$ are two rotation matrices, each a function of three independent coordinates such as the Euler angles, and where the $a_j$, $b_j$ and $c_j$ are twelve constants forming three linearly independent 4-vectors $\bf a$, $\bf b$ and $\bf c$, in the mass space, orthogonal to the mass 4-vector ${\bf m}  = ( m_1 , m_2 , m_3 , m_4)$:
\begin{equation} 
\begin{array}{c}
a_1 m_1 + a_2 m_2 + a_3 m_3 + a_4 m_4 = 0 \, ,\\
b_1 m_1 + b_2 m_2 + b_3 m_3 + b_4 m_4 = 0 \, ,\\
c_1 m_1 + c_2 m_2 + c_3 m_3 + c_4 m_4 = 0 \, .
\end{array}\, , \label{base}
\end{equation}

We introduce the following notation for the matrix
\begin{equation}
{\bf M} = \left( \begin{array}{cccc}
m_1 & 0 & 0 & 0 \\
0 & m_2 & 0 & 0 \\
0 & 0 & m_3 & 0 \\
0 & 0 & 0 & m_4
\end{array} \right) \, .
\end{equation}

In order to complete the definition of vectors $\bf a$, $\bf b$ and $\bf c$ we assume
\begin{equation} 
{\bf a\, M\, b}^{\rm T} = 0 \, , \quad {\bf b\, M\, \bf c}^{\rm T} = 0 \, , \quad {\bf c\, M\, a}^{\rm T} = 0 \, , \label{ort}
\end{equation}

\begin{equation} 
{\bf a\, M}^2\, {\bf b}^{\rm T} = 0 \, , \quad {\bf b\, M}^2\, {\bf c}^{\rm T} = 0 \, , \quad {\bf c\, M}^2\, {\bf a}^{\rm T} = 0 \, \label{equil}
\end{equation}
which determine the directions of $\bf a$, $\bf b$ and $\bf c$ in the 3-plane orthogonal to $\bf m$, and we assume the normalizations
\begin{equation} 
{\bf a\, M\, a}^{\rm T} = {\bf b\, M\, b}^{\rm T} = {\bf c\, M\, c}^{\rm T} = \mu \, , \label{nor}
\end{equation}
that make vectors $\bf a$, $\bf b$ and $\bf c$ adimensional.

These three vectors are easily computed from the previous
orthogonality conditions. One has
\begin{equation}
{\bf a} = \mu y_a \left( \frac{1}{m_1 - x_a}, \frac{1}{m_2 -x_a}, \frac{1}{m_3 - x_a}, \frac{1}{m_4 - x_a} \right) \, ,
\end{equation}
\begin{equation}
{\bf b} =\mu y_b \left( \frac{1}{m_1 - x_b}, \frac{1}{m_2 -x_b}, \frac{1}{m_3 - x_b}, \frac{1}{m_4 - x_b} \right) \, ,
\end{equation}
and
\begin{equation}
{\bf c} =\mu y_c \left( \frac{1}{m_1 - x_c}, \frac{1}{m_2 -x_c}, \frac{1}{m_3 - x_c}, \frac{1}{m_4 - x_c} \right) \, ,
\end{equation}
where $y_a$, $y_b$ and $y_c$ are normalization factors, and $x_a$, $x_b$ and $x_c$ are the roots of the cubic equation
$$
-x^3 (m_1 + m_2 + m_3 + m_4) + 2 x^2 (m_1 m_2 + m_2 m_3 + m_3 m_1 + m_4 m_1 +m_4 m_2 + m_4 m_3)
$$
\begin{equation}
- 3x (m_2 m_3 m_4 + m_3 m_4 m_1 + m_4 m_1 m_2 + m_1 m_2 m_3) + 4 m_1 m_2 m_3 m_4 = 0 \, . \label{char}
\end{equation}
The symmetric nature of this equation is the consequence that this cubic polynomial is related to the derivative of the polynomial $(y-\frac{1}{m_1})(y-\frac{1}{m_2})(y-\frac{1}{m_3})(y-\frac{1}{m_4})$. The roots of this derivative are: $1/x_a$, $1/x_b$, $1/x_c$, and are located between the inverses of the masses.

These quantities are defined in this form only for different masses. In that case we have the inequalities
\begin{equation}
m_1 > x_a > m_2 > x_b > m_3 > x_c > m_4\, ,
\end{equation}
that imply
\begin{equation}
\begin{array}{c}
a_1>0, a_2<0, a_3<0, a_4<0\, ; \\
b_1>0, b_2>0, b_3<0, b_4<0\, ; \\
c_1>0, c_2>0, c_3>0, c_4<0\, .
\end{array}
\end{equation}

The column elements of the constant matrix
\begin{equation}
{\bf E} = \left(\begin{array}{cccc}
a_1 & a_2 & a_3 & a_4 \\
b_1 & b_2 & b_3 & b_4 \\
c_1 & c_2 & c_3 & c_4
\end{array} \right)\, ,
\end{equation}
are the coordinates of the four vertices of a rigid orthogonal tetrahedron.

An orthogonal tetrahedron has the property that the perpendicular lines to the faces trough the four vertices intersect at the same point. Orthogonal tetrahedra were considered by Lagrange in 1773 \cite{la}. Other old references on orthogonal tetrahedra are found in a paper by Court \cite{co}, where he calls it orthocentric. Placing the four masses at the corresponding vertices, that intersection point is actually the center of mass of the four masses, and the moment of inertia tensor of the four particles has the same principal value in any direction. Equations (\ref{ort}) and (\ref{nor}) imply that the inertia tensor of the rigid tetrahedron is proportional by a factor $2 \mu$ to the unit matrix.

To show these properties we consider a four vector linearly independent to the three 4-vectors $\bf a$, $\bf b$ and $\bf c$
\begin{equation}
{\bf d} = \sqrt{\frac{\mu}{m}} (1, 1, 1, 1)\, , \label{cuarto}
\end{equation}
where we use the notation $m = m_1 + m_2 + m_3 + m_4$ for the total mass of the system. Then using definitions (\ref{base}), (\ref{ort}), (\ref{nor}), and (\ref{cuarto}) we write them in terms of $r = \sqrt{\frac{\mu}{m}}$ in the form
\begin{equation}
\frac{1}{\mu} \left( \begin{array}{cccc}
a_1 & a_2 & a_3 & a_4 \\
b_1 & b_2 & b_3 & b_4 \\
c_1 & c_2 & c_3 & c_4 \\
r & r & r & r
\end {array} \right) M \left( \begin{array}{cccc}
a_1 & b_1 & c_1 & r \\
a_2 & b_2 & c_2 & r \\
a_3 & b_3 & c_3 & r \\
a_4 & b_4 & c_4 & r
\end {array} \right) = \left( \begin{array}{cccc}
1 & 0 & 0 & 0 \\
0 & 1 & 0 & 0 \\
0 & 0 & 1 & 0 \\
0 & 0 & 0 & 1
\end {array} \right)\, .
\end{equation}
Since the inverse matrix from the left is equal to the inverse from the right, this equation transforms into
\begin{equation}
\left( \begin{array}{cccc}
a_1 & b_1 & c_1 & r \\
a_2 & b_2 & c_2 & r \\
a_3 & b_3 & c_3 & r \\
a_4 & b_4 & c_4 & r
\end {array} \right) \left( \begin{array}{cccc}
a_1 & a_2 & a_3 & a_4 \\
b_1 & b_2 & b_3 & b_4 \\
c_1 & c_2 & c_3 & c_4 \\
r & r & r & r
\end {array} \right) = \left( \begin{array}{cccc}
\frac{\mu}{m_1} & 0 & 0 & 0 \\
0 & \frac{\mu}{m_2} & 0 & 0 \\
0 & 0 & \frac{\mu}{m_3} & 0 \\
0 & 0 & 0 & \frac{\mu}{m_4}
\end {array} \right)\, .
\end{equation}
Because this matrix equation is equal to its transposed; it just has ten independent equations. Four of them are
\begin{equation}
a_j^2 + b_j^2 + c_j^2 = \mu \left( \frac{1}{m_j} - \frac{1}{m}\right)\, , \quad (j=1, 2, 3, 4) \label{20}
\end{equation}
The other six are
\begin{equation}
a_i a_j + b_i b_j + c_i c_j = - \frac{\mu}{m}\, . \mbox{ ($i \neq j$ )} \label{21}
\end{equation}
From these basic equations it is easy to show that the position vector of one vertex is orthogonal to the three vectors between two vertices of the corresponding face (to the first vertex.)
\begin{equation}
a_i (a_j - a_k) + b_i (b_j - b_k) + c_i (c_j - c_k) = 0 \quad \mbox{ ($i,j,k$ different) }\, .
\end{equation}
In addition, the distance between two vertices is given by
\begin{equation}
(a_i - a_j)^2 + (b_i - b_j)^2 + (c_i - c_j)^2 = \mu \left( \frac{1}{m_i} + \frac{1}{m_j}\right)\, . \label{lados}
\end{equation}
This is the condition to have a moment of inertia tensor with the same three principal moments of inertia. The six edges of the tetrahedron should be equal (proportional) to the square root of the right hand side of this equation. The volume of this tetrahedron is equal to 1/6.

There are other remarkable geometrical properties of an orthogonal tetrahedron. The center of mass of each face is at the orthocenter where the three altitudes of the face intersect. This point is on the same straight line between the opposite vertex and the center of mass. In addition to the orthogonality of the three sets of two opposite edges of the tetrahedron, the two orthogonal edges are also orthogonal to the line joining the center of mass of the two edges.

Let me do in this paragraph a technical digression that is specially relevant for engineering and physical minds. In formulating the explicit expressions of the coordinates of the constant rigid tetrahedra $E$ the origin for computing the $\bf G'$ rotation was arbitrarily chosen to be the one associated with the equilateral tetrahedra with four different masses, which implies a constant $\bf G'$ that was here selected equal to the unit matrix. This convention is introduced through equations (\ref{equil}) that are actually not necessary for the rest of the statements and proofs in this paper. Although there are of course other important coordinate systems to fix the origin for measuring the $\bf G'$ rotation; from these I prefer to choose one particle along one coordinate axis, the other three in a parallel plane to the parallel coordinate plane which does not include the  first particle; a second particle on an orthogonal coordinate plane that includes the first particle, and the other two particles on a line that is parallel to a coordinate axis and perpendicular to the coordinate plane of the first two particles.
Another equally important referential for the origin of the rigid tetrahedron is associated with the grouping of the four particles in two sets of two particles. The center of mass of the two pairs, and the center of mass for the whole system are on a coordinate axis, and each of the two selected pairs of particles are placed on a line parallel to a coordinate axis.

The previous definitions do not work in the important cases when two or more masses have exactly the same value. In those cases the tetrahedron is identified more easily from condition (\ref{lados}) in terms of the masses. The selection of the origin for measure the rotation $\bf G'$ is now forced by the symmetry of the tetrahedron introduced by the mass equality.

This rigid tetrahedron is the generalization of the rigid triangle of the Three-Body problem with the center of mass at the orthocenter discussed previously in \cite{pb}.

I assume for simplicity that the potential energy is given by the Newton potential (the gravitational constant is equal to 1)
\begin{equation}
V = - \sum_{i<j}^3\frac{m_i m_j}{r_{ij}}\, ,
\end{equation}
although our results may be generalized for any potential which is a given power law of the relative distances between particles $r_{ij}$. It follows the relation between the interparticle distance and the new coordinates. The relative position between particles $i$ and $j$ is
\begin{equation}
\left( \begin{array}{c}
x_j - x_i \\
y_j - y_i \\
z_j - z_i
\end{array} \right) = {\bf G} \left( \begin{array}{ccc}
R_1 & 0 & 0 \\
0 & R_2 & 0 \\
0 & 0 & R_3
\end{array} \right)  {\bf G'}^{\rm T} \left( \begin{array}{c}
a_j - a_i \\
b_j - b_i \\
c_j - c_i
\end{array} \right)\, .
\end{equation}
The square of this vector is not a function of the first rotation $\bf G$, but just of the scale matrix and the second rotation matrix
\begin{equation}
r_{ij}^2 = (a_j - a_i \ b_j - b_i \ c_j - c_i) {\bf A} \left( \begin{array}{c}
a_j - a_i \\
b_j - b_i \\
c_j - c_i
\end{array} \right)\, .
\end{equation}

where $\bf A$ is the symmetric matrix
\begin{equation} 
{\bf A}  = \left( \begin{array}{ccc}
A_{11} & A_{12} & A_{13} \\
A_{12} & A_{22} & A_{23} \\
A_{13} & A_{23} & A_{33}
\end{array} \right) = {\bf G'} \left( \begin{array}{ccc}
R_1^2 & 0 & 0 \\
0 & R_2^2 & 0 \\
0 & 0 & R_3^2
\end{array} \right) {\bf G'}^{\rm T} \, .
\end{equation}
The six distances are thus functions of six components of matrix $A$ or equivalently, are functions of the six independent coordinates in the scales $R_i$, and the rotation $\bf G'$.

We also compute the kinetic energy as a function of the new coordinates, which is given by
$$
K = {\mu \over 2} \left[ \sum_{i=1}^3 \dot{R_i}^2 - 4( R_2 R_3 \omega_1 \Omega_1 + R_3 R_1 \omega_2 \Omega_2 + R_1 R_2 \omega_3 \Omega_3) \right. +
$$
$$
\omega^{\rm T} \left(\begin{array}{ccc}
R_2^2 + R_3^2 & 0 & 0 \\
0 & R_3^2 + R_1^2 & 0 \\
0 & 0 & R_1^2 + R_2^2
\end{array} \right) \omega +
$$
\begin{equation}
\left.\Omega^{\rm T} \left(\begin{array}{ccc}
R_2^2 + R_3^2 & 0 & 0 \\
0 & R_3^2 + R_1^2 & 0 \\
0 & 0 & R_1^2 + R_2^2
\end{array} \right) \Omega  \right]\, ,
\end{equation}
where $\omega = (\omega_1, \omega_2, \omega_3)$ is the angular velocity vector of the first rotation $\bf G$, and $\Omega = (\Omega_1, \Omega_2, \Omega_3)$ is the corresponding angular velocity vector of the second rotation $\bf G'$.

\section{Equations of Motion}
The equations of motion follow from the Lagrange equations derived from the Lagrangian $K - V$ as presented in any standard text on Mechanics \cite{ll}, \cite{js}.

However, the three coordinates related to the first rotation produces Lagrange equations that imply, when the potential energy is a function only of the distances, conservation of the angular momentum vector in the inertial system
\begin{equation}
{\bf G} \; \, {\bf L}
\end{equation}
where $\bf L$ is the angular momentum in the principal moments of inertia frame
$$
{\bf L} = \frac{\partial K}{\partial \omega}
$$
\begin{equation}
= \mu \left(\begin{array}{c}
(R_2^2 + R_3^2) \omega_1 \\
(R_3^2 + R_1^2) \omega_2 \\
(R_1^2 + R_2^2) \omega_3
\end{array} \right) - 2 \mu \left(\begin{array}{c}
R_2 R_3 \Omega_1 \\
R_3 R_1 \Omega_2 \\
R_1 R_2 \Omega_3
\end{array} \right)
\end{equation}

This conservation leads to three first order equations forming, for this four body problem a generalization of the Euler equations valid for the rotation of a rigid body, namely
$$
\frac{d}{d t} \left( \begin{array}{c}
\mu (R_2^2 + R_3^2) \omega_1 - 2 \mu R_2 R_3 \Omega_1 \\
\mu (R_3^2 + R_1^2) \omega_2 - 2 \mu R_3 R_1 \Omega_2 \\
\mu (R_1^2 + R_2^2) \omega_3 - 2 \mu R_1 R_2 \Omega_3
\end{array} \right) =
$$
\begin{equation}
\left( \begin{array}{ccc}
\mu (R_3^2 - R_1^2) \omega_2 \omega_3 + 2 \mu R_1 (R_2 \omega_2 \Omega_3 - R_3 \omega_3 \Omega_2) \\
\mu (R_1^2 - R_2^2) \omega_3 \omega_1 + 2 \mu R_2 (R_3 \omega_3 \Omega_1 - R_1 \omega_1 \Omega_3) \\
\mu (R_2^2 - R_3^2) \omega_1 \omega_2 + 2 \mu R_3 (R_1 \omega_1 \Omega_2 - R_2 \omega_2 \Omega_1)
\end{array}\right)\, . \label{euler}
\end{equation}

The so called \textit{elimination of the nodes} in the Three-Body problem \cite{wh}, has a similar representation in this coordinates for the Four-Body problem by means of the equation that equals the angular momentum vector in the principal moments of inertia frame to the rotation of a constant vector, which may be written in terms of two Euler angles
\begin{equation}
\mu \left(\begin{array}{c}
(R_2^2 + R_3^2) \omega_1 \\
(R_3^2 + R_1^2) \omega_2 \\
(R_1^2 + R_2^2) \omega_3
\end{array} \right) - 2 \mu \left(\begin{array}{c}
R_2 R_3 \Omega_1 \\
R_3 R_1 \Omega_2 \\
R_1 R_2 \Omega_3
\end{array} \right) = \ell {\bf G}^{\rm T} \left( \begin{array}{c}
0 \\
0 \\
1
\end{array} \right)\, ,
\end{equation}
where $\ell$ is the magnitude of the conserved angular momentum.

The Lagrangian equations of motion for the three scale coordinates are
\begin{equation} 
\mu \frac{d^2}{d t^2} R_1 + 2 \mu [R_2 \omega_3 \Omega_3 + R_3 \omega_2 \Omega_2 ] + \mu R_1 (\omega_2^2 + \omega_3^2 +\Omega_2^2 + \Omega_3^2) = - \frac{\partial V}{\partial R_1} \, ,
\end{equation}
\begin{equation}
\mu \frac{d^2}{d t^2} R_2 + 2 \mu [R_3 \omega_1 \Omega_1 + R_1 \omega_3 \Omega_3 ] + \mu R_2 (\omega_3^2 + \omega_1^2 +\Omega_3^2 + \Omega_1^2) = - \frac{\partial V}{\partial R_2} \, ,
\end{equation}
and
\begin{equation}
\mu \frac{d^2}{d t^2} R_3 + 2 \mu [R_1 \omega_2 \Omega_2 + R_2 \omega_1 \Omega_1 ] + \mu R_3 (\omega_1^2 + \omega_2^2 + \Omega_1^2 + \Omega_2^2) = - \frac{\partial V}{\partial R_3} \, .
\end{equation}

The three equations of motion for the three coordinates associated with the second rotation $\bf G'$ are written as an Euler equation similar to the one found for the first rotation, although the internal angular momentum is not conserved because of the presence of an internal torque
$$
\frac{d}{d t} \left( \begin{array}{c}
\mu (R_2^2 + R_3^2) \Omega_1 - 2 \mu R_2 R_3 \omega_1 \\
\mu (R_3^2 + R_1^2) \Omega_2 - 2 \mu R_3 R_1 \omega_2 \\
\mu (R_1^2 + R_2^2) \Omega_3 - 2 \mu R_1 R_2 \omega_3
\end{array} \right) = \left( \begin{array}{c}
K_1 \\
K_2 \\
K_3
\end{array} \right) +
$$
\begin{equation}
\left( \begin{array}{ccc}
\mu (R_3^2 - R_1^2) \Omega_2 \Omega_3 - 2 \mu R_1 (R_2 \omega_2 \Omega_3 - R_3 \omega_3 \Omega_2) \\
\mu (R_1^2 - R_2^2) \Omega_3 \Omega_1 - 2 \mu R_2 (R_3 \omega_3 \Omega_1 - R_1 \omega_1 \Omega_3) \\
\mu (R_2^2 - R_3^2) \Omega_1 \Omega_2 - 2 \mu R_3 (R_1 \omega_1 \Omega_2 - R_2 \omega_2 \Omega_1)
\end{array}\right)\, ,
\end{equation}
where $K_1, K_2, K_3$ are the components of the internal torque $\bf K$ which is expressed in terms of the derivatives of the potential energy with respect to the three independent coordinates $q_j$ in the rotation $\bf G'$ and the three vectors ${\bf c}_j$ that appear in the expression of the angular velocity in terms of the same coordinates
\begin{equation}
\Omega = \sum_{j=1}^3 {\bf c}_j \dot{q}_j
\end{equation}
where the vectors ${\bf c}_j$ are generally functions of the coordinates $q_j$.

The internal torque is determined by the equations
\begin{equation}
{\bf K} \cdot {\bf c}_j = \frac{\partial V}{\partial q_j}\, .
\end{equation}

There is one  more constant of motion, namely the total energy
$$
E = V + K = V + {\mu \over 2} \left[ \sum_{i=1}^3 \dot{R_i}^2 - 4 ( R_2 R_3 \omega_1 \Omega_1 + R_3 R_1 \omega_2 \Omega_2 + R_1 R_2 \omega_3 \Omega_3) \right. +
$$
$$
\omega^{\rm T} \left(\begin{array}{ccc}
R_2^2 + R_3^2 & 0 & 0 \\
0 & R_3^2 + R_1^2 & 0 \\
0 & 0 & R_1^2 + R_2^2
\end{array} \right) \omega +
$$
\begin{equation}
\left.\Omega^{\rm T} \left(\begin{array}{ccc}
R_2^2 + R_3^2 & 0 & 0 \\
0 & R_3^2 + R_1^2 & 0 \\
0 & 0 & R_1^2 + R_2^2
\end{array} \right) \Omega  \right]\, .
\end{equation}

 \section{The plane problem}
The case with the four particles in a constant plane is an important and old subject \cite{dz}. Our coordinates are now adapted to that case. The third component of the cartesian coordinates of the four particles are zero. The modification of our coordinates (4) for this case is given by two changes: the first rotation by just one angle in the plane of motion; and the scale associated with the third coordinate is zero, namely
$$
\left(\begin{array}{cccc}
x_1 & x_2 & x_3 & x_4 \\
y_1 & y_2 & y_3 & y_4 \\
0 & 0 & 0 & 0
\end{array} \right) =
$$
\begin{equation}
\left(\begin{array}{ccc}
\cos \psi & -\sin \psi & 0\\
\sin \psi & \cos \psi & 0 \\
0 & 0 & 1
\end{array} \right)
 \left(\begin{array}{ccc}
R_1 & 0 & 0 \\
0 & R_2 & 0 \\
0 & 0 & 0
\end{array} \right) {\bf G'}^{\rm T}
\left(\begin{array}{cccc}
a_1 & a_2 & a_3 & a_4 \\
b_1 & b_2 & b_3 & b_4 \\
c_1 & c_2 & c_3 & c_4
\end{array} \right)\, .
\end{equation}
This equation simplifies to
$$
\left(\begin{array}{cccc}
x_1 & x_2 & x_3 & x_4 \\
y_1 & y_2 & y_3 & y_4
\end{array} \right) =
$$
\begin{equation}
\left(\begin{array}{cc}
\cos \psi & -\sin \psi \\
\sin \psi & \cos \psi
\end{array} \right) \left(\begin{array}{ccc}
R_1 & 0 & 0 \\
0 & R_2 & 0
\end{array} \right) {\bf G'}^{\rm T}
\left(\begin{array}{cccc}
a_1 & a_2 & a_3 & a_4 \\
b_1 & b_2 & b_3 & b_4 \\
c_1 & c_2 & c_3 & c_4
\end{array} \right)\, , \label{flat}
\end{equation}
in terms of six degrees of freedom.

We need three independent coordinates (for example three Euler angles) in $\bf G'$ for the two independent vectors in four dimensions expressed in the base of the three constant vectors $\bf a$, $\bf b$, and $\bf c$, orthogonal to the mass vector.

Let me do in this paragraph a technical digression that is specially interesting for engineering or physical minds. We must formulate in a mathematical language the conditions for a plane solution. The most usual way to do this is to equal to zero the Cayley-Menger determinant which has entries equal to 1, 0, and the squares of the distances between particles. Although Dziobek  \cite{dz} considered this approach of paramount importance, however he introduced equivalent conditions that have been promoted by many years by A. Albouy and coworkers (see \cite{al} and references therein,) which consists in using the four directed areas of the triangles formed by the particles.

The four (twice) directed areas are written in terms of the cartesian coordinates as
$$
S_1 = \left| \begin{array}{ccc}
1 & 1 & 1 \\
x_2 & x_3 & x_4 \\
y_2 & y_3 & y_4
\end{array} \right| \, , \quad S_2 = \left| \begin{array}{ccc}
1 & 1 & 1 \\
x_1 & x_4 & x_3 \\
y_1 & y_4 & y_3
\end{array} \right| \, ,
$$
\begin{equation}
S_3 = \left| \begin{array}{ccc}
1 & 1 & 1 \\
x_1 & x_2 & x_4 \\
y_1 & y_2 & y_4
\end{array} \right| \, , \quad S_4 = \left| \begin{array}{ccc}
1 & 1 & 1 \\
x_1 & x_3 & x_2 \\
y_1 & y_3 & y_2
\end{array} \right| \, ,
\end{equation}
that are the four signed 3 $\times$ 3 minors formed from the matrix
\begin{equation}
\left(\begin{array}{cccc}
1 & 1 & 1 & 1 \\
x_1 & x_2 & x_3 & x_4 \\
y_1 & y_2 & y_3 & y_4
\end{array} \right)
\end{equation}

Addition to the previous matrix of a row equal to any of its three rows produces a square matrix with determinant zero, that implies that the necessary and sufficient conditions to have a constant plane tetrahedron are
\begin{equation}
\sum_{i=1}^4 S_i = 0\, ,\label{plan}
\end{equation}
and
\begin{equation}
\sum_{i=1}^4 S_i x_i = 0\, , \quad \sum_{i=1}^4 S_i y_i = 0\, .
\end{equation}
The two last equations summarized by the zero vector condition
\begin{equation}
\sum_{i=1}^4 S_i {\bf r}_i = {\bf 0}\, . \label{afin}
\end{equation}

An expression for the three directed areas in terms of the previous coordinates follows
\begin{equation}
\left( \begin{array}{c}
S_1 \\
S_2 \\
S_3 \\
S_4
\end{array} \right) =  C M E^{\rm T} {\bf G'} \left( \begin{array}{c}
0 \\
0 \\
1
\end{array} \right)\, , \label{areas}
\end{equation}
where $C$ is a constant with units of area over mass. Substitution of equations (\ref{flat}) and (\ref{areas}) in equations (\ref{plan}) or (\ref{afin}) one obtains an identity, independent of coordinates $R_1$, $R_2$, $\psi$ and the rotation angle of $\bf G'$ around the unit vector
\begin{equation}
{\bf G'} \left( \begin{array}{c}
0 \\
0 \\
1
\end{array} \right)\, .
\end{equation}
Given the four masses, the four directed areas of the four particles are functions of this unit vector direction only, up to a multiplicative constant $C$ depending on the choice of physical units. These explicit expressions should make clear
 Albouy's \cite{al} affine formulation of the plane condition. Another form of this constant plane condition is also published in reference \cite {pl}.

In the plane case the angular momentum has a constant direction orthogonal to the plane and of magnitude
\begin{equation}
P_\psi = \frac{\partial K}{\partial \dot{\psi}} = \mu [\dot{\psi}(R_1^2 + R_2^2) - 2 R_1 R_2 \Omega_3]\, . \label{mom}
\end{equation}
The kinetic energy becomes
$$
K = {\mu \over 2} \left[ \sum_{i=1}^2 \dot{R_i}^2 - 4( R_1 R_2 \dot{\psi} \Omega_3)+ \dot{\psi}^2 (R_1^2 + R_2^2) \right.+
$$
\begin{equation}
\left.\Omega^{\rm T} \left(\begin{array}{ccc}
R_2^2 & 0 & 0 \\
0 & R_1^2 & 0 \\
0 & 0 & R_1^2 + R_2^2
\end{array} \right) \Omega  \right]\, ,
\end{equation}

Substitution of polar coordinates for the $R_1$ and $R_2$ coordinates
\begin{equation}
R_1 = R \cos \theta \, , \quad R_2 = R \sin \theta
\end{equation}
and writing the kinetic energy in terms of the angular momentum constant of motion instead of the $\dot{\psi}$ velocity lead us to
\begin{equation}
K = {\mu \over 2} \left[ \dot{R}^2 + R^2 \left(\dot{\theta}^2 + \Omega_3^2 \cos^2 (2 \theta) + \Omega_1^2 \sin^2 \theta + \Omega_2^2 \cos^2 \theta \right) \right] +\frac{P_\psi^2}{2 \mu R^2}\, .
\end{equation}
Energy conservation is thus expressed as
\begin{equation}%
E = {\mu \over 2} \left[ \dot{R}^2 + R^2 \left(\dot{\theta}^2 + \Omega_3^2 \cos^2 (2 \theta) + \Omega_1^2 \sin^2 \theta + \Omega_2^2 \cos^2 \theta \right) \right] + \frac{P_\psi^2}{2 \mu R^2} + V \, , \label{ene}
\end{equation}
where $V$ represents the potential energy.

\section{Central configurations}
In this section we begin with the approach by Dziobek \cite{dz}. See reference \cite{sa} for a contemporary approach. The Four-Body central configurations are determined as critical points of the potential energy with a fixed total inertia moment that in three dimensional space lead to
\begin{equation}
m_i m_j /r_{ij}^3 = \sigma m_i m_j  \label{var}\, .
\end{equation}
The left hand side of this equation is the derivative of the potential energy with respect to $r_{ij}^2$. The right hand side is the derivative of the inertia moment with respect to $r_{ij}^2$ multiplied by an unknown constant $\sigma$ that includes the constant total mass, contained in the expression for the total inertia moment, and the gravity constant in the potential function. Of course, this equation simplifies to
\begin{equation}
1/r_{ij}^3 = \sigma \, . \label{three}
\end{equation}

According to equation (\ref{three}), the only Four-Body three dimensional central configuration results just if the six distances are the same, giving an equilateral tetrahedron. For an equilateral tetrahedron one particular coordinate system is given placing its vertices on alternating corners of a cube having the six faces normal to the coordinate axis. Then the center of mass is computed and the origin of coordinates translated to this position. After that the tensor of inertia is determined and the scale factors associated to the principal moments of inertia are in the ratios
\begin{equation}\
R_1^2/ R_2^2/ R_3^2 = x_a/ x_b/ x_c\, .
\end{equation}
This fact is deduced because the $R_j^2$ obey the same characteristic equation (\ref{char}). Now we show that the equilateral tetrahedron has been selected as origin for measuring the ${\bf G'}$ rotation. A tetrahedron with positions
\begin{equation}
\left( \begin{array}{ccc}
\sqrt{x_a/\mu} & 0 & 0 \\
0 & \sqrt{x_b/\mu} & 0 \\
0 & 0 & \sqrt{x_c/\mu}
\end{array} \right) \left(\begin{array}{cccc}
a_1 & a_2 & a_3 & a_4 \\
b_1 & b_2 & b_3 & b_4 \\
c_1 & c_2 & c_3 & c_4
\end{array} \right)
\end{equation}
is an equilateral tetrahedron (namely the ${\bf G'}$ rotation is the unit matrix for the equilateral case). The proof is obtained from this equation by direct computation of the length of the edges and use of equations (\ref{20}) and (\ref{21}). The six edges of that tetrahedron are equal to $\sqrt{2}$.

The equilateral tetrahedron gives a well known solution in which the masses move on straight lines. This is a theorem by Laplace \cite {bo} with an old history.

The non collinear plane central configurations are characterized in our coordinates by constant values of the $\bf G'$ matrix and of the coordinate $\theta$ associated to the constant value of the ratio $R_1/R_2$. For these cases the angular velocity vector $\Omega$ is the null vector, the angular velocity $\dot{\theta}$ is also zero and the equations for conservation of moment and energy, (\ref{mom}) and (\ref{ene}) respectively, become
\begin{equation}%
P_\psi = \mu \dot{\psi} R^2 \, .
\end{equation}
and
\begin{equation}%
E = {\mu \over 2} \dot{R}^2 +\frac{P_\psi^2}{2 \mu R^2} + V \, .
\end{equation}

These equations are identical to similar equations obtained for the Euler and Lagrange central configurations of the Three-Body problem \cite{bp}. They are formally the same as the equations for the conics in the Two-Body problem in terms of the radius $R$ and the true anomaly $\psi$.

 The constant values of the $\bf G'$ matrix and angle $\theta$ refered to above are not arbitrary but they are determined by three independent quantities as discussed in the following.

The plane solutions with zero volume but finite area are obtained taking in account that the variational equation (\ref{var}) is modified adding the restriction of plane motion. This condition is obtained by Dionzek \cite{dz} from the derivative of the Cayley-Menzer determinant with respect to $r_{ij}^2$ that he found to be proportional to the product of the directed areas $S_i S_j$.

It follows that the solution is given in terms of parameters $\lambda$ and $\sigma$
\begin{equation}
r_{jk}^{-3} = \sigma + \lambda A_j A_k\, ,
\end{equation}
where $A_j = S_j/m_j$ are weighted areas, quotient of the directed area divided by the corresponding mass. This equation was presented by Dziobek \cite{dz}. A proof was published by Moeckel \cite{mo}, and using a different approach to the same problem, deduced by Albouy \cite{al}. A new proof of the equation was obtained in a different approach by Pi\~na and Lonngi \cite{pl}.

It follows from (\ref{areas}) that in a plane solution the weighted directed areas are expressed as
\begin{equation}
\left( \begin{array}{c}
A_1 \\
A_2 \\
A_3 \\
A_4
\end{array} \right) =  C E^{\rm T} {\bf G'} \left( \begin{array}{c}
0 \\
0 \\
1
\end{array} \right)\, . \label{sec}
\end{equation}
The weighted directed areas are up to a normalization factor equal to the third rotated coordinate of the rigid tetrahedra.

The weighted directed areas obey the condition
\begin{equation}
\sum_j A_j m_j = 0
\end{equation}
expressing the fact that the sum of the directed areas is zero, equation (\ref{plan}).

Since the lengths and masses are defined up to arbitrary units, I assume with no loss of generality that the parameter $\sigma$ equals unity.
\begin{equation}
r_{jk}^{-3} = 1 + \lambda A_j A_k\, \quad (j \neq k).
\end{equation}
This equation has been considered giving particular values of the four masses and computing the six distances $r_{jk}$ by solving it under restriction (\ref{plan}). The weighted directed areas $A_j$ are functions of the distances and the masses only. Some examples of such approach are \cite{al}, \cite{ja}, \cite{sh}.  According to D. Saari \cite{sa} the problem with this perspective is difficult to manipulate, but we found it to be perfectly feasible as follows below.

In the paper by Pi\~na and Lonngi \cite{pl} a different point of view was adopted, namely that the directed weighted areas (that are defined with a simple functional dependence with respect to the masses,) are known as four given constants. The previous equation then gives the distances as functions of the unknown parameter $\lambda$. Trough them, the areas of the four triangles become functions of $\lambda$, that should obey the necessary restriction (\ref{plan}), to verify that one has a plane solution. This restriction allows in many cases to determine the value of $\lambda$ and hence the values of the six distances and the four masses. This is an implicit way to deduce planar central configurations with four masses.

\begin{figure}
\centering
\scalebox{0.5}{\includegraphics{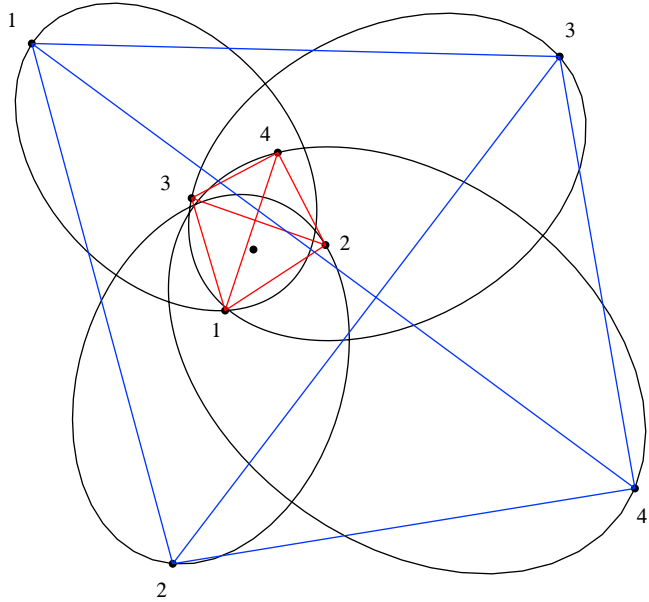}}

\caption{Two different sets of simultaneous positions of four particles with different masses following elliptic trajectories in a central convex plane configuration. The isolated point is at the center of mass at the common focus of the four ellipses. The eccentricity of the four ellipses is $e=0.72$. }
\end{figure}

From the distances and masses one determines the positions of the four particles in the plane frame of principal moments of inertia and the principal moments of inertia are also computed. This allows to know eight components of the rotated rigid tetrahedron $E {\bf G'}$ and the remaining coordinates are known from the four given weighted areas constants $A_j$ according to equation (\ref{sec})

Following this method we computed the necessary data to plot the trajectories of the four particles with different masses represented in Fig. 1. We started from the four constants
$$
A_1 = 6\, , \quad A_2 = -15 \, , \quad A_3 = 4 \, , \quad A_4 = -3
$$
The constant plane conditions give us the value
$$
\lambda = - 0.01268487093192263...
$$
from which the distances are
$$
\begin{array}{l}
r_{23} = 1.12863753386515...\\
r_{31} = 0.933745641175193...\\
r_{12} = 0.953868245971217...\\
r_{41} = 1.325724 35746881...\\
r_{42} = 0.828080639336103...\\
r_{43} = 0.775802698722361...
\end{array}
$$
and the corresponding masses
$$
\begin{array}{l}
m_1 = 0.428260218865972...\\
m_2 = 0.355184464717379...\\
m_3 = 0.261905866491155...\\
m_4 = 0.113826160081235...
\end{array}
$$

This information is sufficient to compute an initial central configuration and from it the four constant coordinates determining $\bf G'$ and $\theta$. The elliptic choreography (with no symmetry) is obtained simply by writing coordinate $R$ as a function of the real anomaly $\psi$. The {\sl latus rectum}, the eccentricity and the initial value of $\psi$ may all be selected arbitrarily.

\end{document}